\documentclass[aps, preprint, showpacs, epsfig]{revtex4}
\usepackage{graphics}
\usepackage{graphicx}
\usepackage{epsfig}

\def \be {\begin{equation}}
\def \ee {\end{equation}}
\def \ba {\begin{eqnarray}}
\def \ea {\end{eqnarray}}
\def \bm {\begin{displaymath}}
\def \em {\end{displaymath}}

\begin{document}
\title{Pair correlation functions in nematics, free-energy functional and isotropic-nematic transition}
\author{Pankaj Mishra and Yashwant Singh}
\author{}
\affiliation{Department of Physics, Banaras Hindu University, 
Varanasi-221 005,
India}
\date{\today}
\begin{abstract}
We develop a free energy functional for an inhomogeneous system that contains both symmetry 
conserved and symmetry broken parts of the direct pair correlation function. These correlation 
functions are found by solving the Ornstein- Zernike equation with the Percus-Yevick closure relation. 
The method developed here gives the pair correlation functions in the ordered phase with features that 
agree well with the results found by computer simulations. The theory predicts accurately
the isotropic-nematic transition in a system of anisotropic molecules and can easily be extended 
to study other ordered phases such as smectics and crystalline solids.
\end{abstract}
\maketitle
The freezing of a fluid of anisotropic molecules into a uniaxial nematic phase
is a typical example of a first-order phase transition in which the continuous symmetry of the
isotropic phase is broken[1]. In a nematic phase, molecules are
aligned along a particular but arbitrary direction so as to have a long range order in orientation
while the translational degrees of freedom remain disordered as in the isotropic fluid. At the 
isotropic-nematic transition the isotropy of the space is spontaneously broken and,
as a consequence, the correlations in the distribution of molecules loose their rotational invariance. By 
computer simulations of a system of ellipsoids Phoung and Schmid [2] have recently evaluated
the effect of breaking of rotational symmetry on pair correlation functions (PCFs) and
showed that in the nematic phase there are two qualitatively different contributions
: one that preserves rotational invariance
and other that breaks it and vanishes in the isotropic phase. The symmetry preserving
part of the PCFs passes smoothly without any abrupt change through the isotropic-nematic
transition. This we expect to be true in most of the symmetry breaking first-order
transitions.

The correlation functions that describe the distribution of molecules in a classical
fluid can be given as the simultaneous solution of an integral equation, the
Ornstein Zernike (OZ) equation, and an algebraic closure relation that relates
correlation functions to the pair potential. Well known approximations to the 
closure relation are the Percus-Yevick (PY) relation, the hypernetted-chain (HNC)
relation and the mean spherical approximation (MSA)[3]. This approach has been used 
quite successfully to describe the structure of isotropic fluids. However,
the application of the theory to ordered phases which can be regarded as an
inhomogeneous "fluid", has so for been very limited though no features of the
theory inherently prevent it from being used to describe structures of the 
ordered phases like crystalline solids, liquid crystals, etc. Holovko and
Sokolovska [4] have used the MSA and Lovett equation[5] which relates 
one-particle density to the PCFs to solve analytically the OZ equation
for a model of spherical particles with the long range anisotropic 
interaction and determined the PCFs in the nematic phase. However, when
Phoung and Schmid[2] used the PY closure relation and the Lovett equation
and solved the OZ equation numerically for system of soft ellipsoids nematic 
phase was not found and for this the PY closure was blamed. Here we adopt a method
based on density-functional formalism and show that though the PY relation
may not be very accurate it, however, leads to nematic phase with PCFs harmonic 
coefficients having features similar to those found by simulations[2] and
by analytical solution[4].

A density-functional theory (DFT) of freezing transition requires
an expression of the grand thermodynamic potential of the system
in terms of one- and two particle distribution functions and a relation
that relates the one-particle density distribution $\rho ({\bf x})$ to the PCFs. 
Such a relation is found by minimizing the grand thermodynamic potential 
with respect to $\rho ({\bf x})$ with appropriate constraints[6]. The correlation
functions that appear in these equations are of the ordered phase and are 
functional of $\rho ({\bf x})$. The free energy functional that exist in the
literature[7, 8] and have been used to study the freezing transitions and other
properties of the ordered phases replace these correlations by that of
the isotropic fluids (see ref.[6] for details). This approximation limits
the applicability of the theory and needs improvement.  

Though in this letter we consider the nematic phase, the method developed here 
is equally applicable to other phases such as smectics and crystalline solids. 
We first show how the fact that the PCFs in the ordered phase have two
distinct parts, leads us to divide the OZ equation and the closure relation into
two sets of equations. The solutions of these equations give us both the symmetry conserving
and symmetry breaking parts of the PCFs.  
Using these correlation functions we construct a free energy functional. 
This free energy functional is then used to calculate the transition parameters
of the isotropic-nematic transition in a model system of elongated rigid molecules 
interacting via the Gay-Berne (GB) pair potential[9].
The GB potential between a pair of molecules $(i,j)$ is written as 
\begin{equation}
u({\bf \hat{r}_{ij}},{\hat {\bf e_i}}, {\hat {\bf e_j}}) =
4 \epsilon({\bf \hat{r}_{ij}},{\hat {\bf e_i}}, {\hat {\bf e_j}})
(R^{-12}-R^{-6})
\end{equation}
where $R=({r_{ij}}-\sigma({\bf \hat{r}_{ij}},{\hat {\bf e_i}}, {\hat {\bf e_j}})+\sigma_0)/\sigma_0$
and ${\bf \hat e}_i$ is the unit vector specifying the axis of symmetry of the $i$th molecule.
The expressions for the angle dependent range parameter $\sigma$ and potential
well depth function $\epsilon$ contain four parameters $x_0, k^{'},\mu $ and $\nu $. 
These parameters measure the anisotropy in the repulsive and attractive
forces. The distance and energy are scaled by parameters $\sigma_0$ and $\epsilon_0$,
respectively. The values of the parameters $x_0, k^{'},\mu $ and $\nu $ are taken 
3.0, 5.0, 2.0 and 1.0 respectively. This model system is found to exhibit
first-order isotropic-nematic transition[10].

The OZ equation which relates the total PCF, $h(\bf{x_1,x_2})$, with 
DPCF, $c(\bf{x_1,x_2})$, in an inhomogeneous system is written as [6,11] 
\be
h({\bf x_1},{\bf x_2})=c({\bf x_1},{\bf x_2})+ 
\int c({\bf x_1},{\bf x_3}) \rho({\bf x_3}) h({\bf x_3},{\bf x_2}) d{\bf x_3}
\ee

where $\bf {x_i}$ indicates both position $\bf r_i$ and orientation ${\bf \Omega_i}$ of the $i$th molecule,
$d{\bf x_3}=d{\bf r_3} d{\bf \Omega_3}$ and $\rho({\bf x_3})$ is the single particle density 
distribution.  The PY relation is expressed as  
\be
c({\bf x_1},{\bf x_2})=(e({\bf x_1},{\bf x_2})-1)[1+h({\bf x_1},{\bf x_2})-c({\bf x_1},{\bf x_2})] 
\ee
where $e(\bf{x_1,x_2})= \exp [-\beta u({\bf x_1},{\bf x_2})]$.
The pair correlation functions $h$ and $c$ are assumed to be sum of two parts; one
which corresponds to rotationally invariant and other which corresponds to the
breaking of the rotational symmetry. Thus
\be
h=h^{(0)}+h^{(n)} {\hspace {0.6cm}}{\rm and}{\hspace {0.6cm}}   c=c^{(0)}+c^{(n)} 
\ee  
This allows us to write the OZ and PY equations as 
\ba
h^{(0)}({\bf x_1},{\bf x_2}) & = & c^{(0)}({\bf x_1},{\bf x_2})+\rho_0 \int c^{(0)}({\bf x_1},{\bf x_3}) \nonumber \\ 
& &\times h^{(0)}({\bf x_3},{\bf x_2}) d{\bf x_3} \\
c^{(0)}({\bf x_1},{\bf x_2}) & = & (e({\bf x_1},{\bf x_2})-1)[1+h^{(0)}({\bf x_1},{\bf x_2})- \nonumber \\
& & c^{(0)}({\bf x_1},{\bf x_2})] 
\ea
 and
\ba
h^{(n)}({\bf x_1},{\bf x_2}) &=& c^{(n)}({\bf x_1},{\bf x_2}) \nonumber \\
&& +\int c^{(0)}({\bf x_1},{\bf x_3})\rho_{n}({\bf x_3}) h^{(0)}({\bf x_3},{\bf x_2}) d{\bf x_3} \nonumber \\
&& +\int (\rho_0+\rho_{n}({\bf x_3})) [c^{(0)}({\bf x_1},{\bf x_3}) h^{(n)}({\bf x_3},{\bf x_2}) \nonumber\\
&& +c^{(n)}({\bf x_1},{\bf x_3}) h^{(0)}({\bf x_3},{\bf x_2}) \nonumber \\
&& +c^{(n)}({\bf x_1},{\bf x_3}) h^{(n)}({\bf x_3},{\bf x_2})]d{\bf x_3} \\ 
c^{(n)}({\bf x_1},{\bf x_2}) & = & (e({\bf x_1},{\bf x_2})-1)[h^{(n)}({\bf x_1},{\bf x_2})- \nonumber \\
&& c^{(n)}({\bf x_1},{\bf x_2})]
\ea 
Here $\rho({\bf x_3})=(\rho_0+\rho_{n}({\bf x_3}))$ where $\rho_0$ is the bulk number density and 
$\rho_n({\bf x_3})= \rho_0 (f({\bf\Omega_3})-1)$. 
$f({\bf\Omega})$ is the single particle orientation distribution function normalized to unity, i.e.
$\int f({\bf\Omega}) d({\bf\Omega})=1$.

Eqs(5) and (6) give relations that are identical to the one used in calculating the PCFs in an 
isotropic phase. Relations given by (7) and (8) are new and as shown below give PCFs arising due to
breaking of symmetry. To solve these equations we chose a coordinate frame where the z-axis points 
in the direction of director $\hat n$ (director frame). All orientation dependent functions are expanded
in spherical harmonics $Y_{lm}(\Omega)$[2,12]. This yields (for uniaxial nematic phase of axially 
symmetric molecules)
\be
f(\Omega) = \frac{1}{\sqrt{4\pi}}\sum_{l({\rm even})}f_{l}Y_{l0}(\Omega) 
\ee
and
\ba
\psi(r,{\Omega_1}, {\Omega_2}) &=& \sum_{l_1 l_2 l m_1 m_2 m}\psi_{l_1 l_2 l m_1 m_2 m}(r)
 Y_{l_1m_1}(\Omega_1) \nonumber \\
&& Y_{l_2m_2}(\Omega_2)Y^{*}_{lm}(\hat r)
\ea 
where $f_l=\sqrt{(2 l + 1)}P_l$ and $\psi$ stands for $h, c$ or $e$. $P_l$ is the order 
parameter; its value is zero in the isotropic phase and non-zero in the nematic phase. 
In uniaxial symmetric phases, only real coefficients
with $m_1+m_2-m=0$ and even $l_1+l_2+l$ enter the expansion. Since the molecules in the model system 
under consideration have axial symmetry, every single $l$ is even as well. Because $h^{(0)}, c^{(0)}$ and
$e$ preserve the rotational symmetry, for them
\be
\psi_{l_1 l_2 l m_1 m_2 m}(r)=\psi_{l_1l_2l}(r) C_g(l_1 l_2 l m_1 m_2 m)
\ee
where $C_g$ is the Clebsch-Gordan coefficient.

We have solved (5) and (6) for the model system of (1) using a method described in ref.[13]
and determined the values of $c^{(0)}_{l_1 l_2 l m_1 m_2 m}(r)$ and 
$h^{(0)}_{l_1 l_2 l m_1 m_2 m}(r)$ for values of $l, l_i$
up to $l_{max}=8$ at reduced temperature $T^*(\equiv k_{B}T/\epsilon_0)=1.0$ and for 
densities $0\le \rho^*(\equiv\rho{\sigma_{0}}^{3})\le 0.36$.

To solve (7)-(8) we first set up linear 
equations for $h^{(n)}_{l_1 l_2 l m_1 m_2 m}(r)$ and  $c^{(n)}_{l_1 l_2 l m_1 m_2 m}(r)$
using the expansions of eqn.(9) and (10).  In these equations 
$h^{(0)}_{l_1 l_2 l m_1 m_2 m}(r)$, $c^{(0)}_{l_1 l_2 l m_1 m_2 m}(r)$ and
the order parameters $P_l$ appear. Here we restrict ourselves to only one order parameter ${P_2}$ and solve these
equations for $0\le P_2 \le 0.70$ at the interval of $\Delta P_2=0.05$ for all densities
between 0 and 0.34 and for $l, l_i$ up to $l_{max}=4$. Though we followed the same iterative method 
as used in the case of isotropic, but examined the behavior of each harmonic coefficients 
$h^{(n)}_{l_1 l_2 l m_1 m_2 m}$ at large distance and ensured that the proper convergence takes place 
(the computational details will be given elsewhere[14]). The PCFs thus generated are used in
constructing the free energy functional.

The reduced free energy $A[\rho]$ of an inhomogeneous system is a functional of density
$\rho({\bf x})$ and is written as [6]
\be
A[\rho]=A_{id}[\rho]+A_{ex}[\rho] 
\ee
The ideal gas part $A_{id}[\rho]$ is exactly known,
\be
A_{id}[\rho]=\int d{\bf x} \rho({\bf x})[\ln \{\rho({\bf x})\Lambda\}-1] 
\ee
where $\Lambda$ is the cube of the thermal wavelength associated with a molecule.
The excess part arising due to intermolecular interaction is related with the 
DPCF of the system as
\ba
\frac{\delta^2 A_{ex}}{\delta \rho({\bf x_1})\delta\rho({\bf x_2})} & = & -c(\bf{x_1, x_2}; [\rho]) \nonumber \\
& = &-c^{(0)}({\bf{x_1, x_2}}; \rho_0) \nonumber \\
&& -c^{(n)}(\bf{x_1, x_2}; [\rho])
\ea
$A_{ex}[\rho]$ is found by functional integration of (14). In this integration the system 
is taken from some initial density to the final density $\rho(\bf{x})$ along a path in the density 
space, the result is independent of the path of the integration[15]. For the symmetry 
conserving part $c^{(0)}$ the
integration in density space is done taking isotropic fluid of density $\rho_l$(the density 
of coexistence fluid) as reference. This leads to 
\ba
A_{ex}^{(0)}[\rho]&=&A_{ex}(\rho_l)-\frac{1}{2}\int d{\bf x_1}\int d{\bf x_2} 
\Delta\rho(\bf {x_1})\Delta\rho({\bf x_2}) \nonumber \\
&& \times {\bar c}({\bf x_1,x_2})
\ea
where 
\ba
{\bar c}({\bf x_1,x_2}) & = & 2\int d\lambda \lambda \int 
d\lambda^{'} c^{(0)}\{{\bf x_1,x_2}; \rho_l+\lambda\lambda^{'}(\rho_0-\rho_l)\} \nonumber  
\ea
 $\Delta\rho({\bf{x}})=\rho({\bf{x}})-\rho_l $, $A_{ex}(\rho_l)$ is the excess reduced free energy of
the isotropic fluid of density $\rho_l$ and $\rho_0$ is the average density of the ordered
phase. 

In order to integrate over $c^{(n)}[\rho]$, we characterize the density 
space by two parameters $\lambda$ and $\xi$ which vary from 0 to 1. The parameter $\lambda$
raises density from $0$ to $\rho_0$ as it varies from 0 to 1 whereas parameter $\xi$  raises 
the order parameter from $0$ to $P_2$ as it varies from 0 to 1. This integration gives 
\be
A_{ex}^{n}[\rho]=-\frac{1}{2}\int d{\bf x_1} \int d{\bf x_2} \rho({\bf x_1}) \rho({\bf x_2}) 
{\tilde c}({\bf x_1,x_2})
\ee
where
\ba
{\tilde c}({\bf x_1,x_2}) & = & 4\int_{0}^{1} d\xi \xi \int_{0}^{1} d\xi^{'} 
\int_{0}^{1} d\lambda \lambda \int_{0}^{1}d\lambda^{'}  \nonumber \\ 
&& \times c^{(n)}({\bf x_1, x_2}, \lambda\lambda^{'}\rho_0; \xi\xi^{'}P_2) \nonumber
\ea
It is important to note that while integrating over $\lambda$  
the order parameter $P_2$ is kept fixed and while integrating over $\xi$ the density
is kept fixed. The result does not depend whether integration is done first over 
$\lambda$ or $\xi$. The free energy functional of an ordered phase is the sum of (13), (15)
and (16). Note that the Ramakrishnan and Yousuff (RY) [7] free energy functional is the sum of only (13) and (15) and 
contains an additional approximation in which ${\bar c}({\bf x_1,x_2})$ in (15) is replaced by 
$c({\bf x_1,x_2}; \rho_l)$. 

The grand thermodynamic potential defined as $-W= A- \beta \mu \int d{\bf x} 
\rho({\bf x})$ where $\mu$ is the chemical potential, 
is preferred to locate the freezing transition as it ensures the pressure and chemical potential of
the two phases remain equal at the transition. The transition point is
determined by the condition $\Delta W= W_{l}- W= 0$. The order parameters are determined
from equations found by minimizing the grand thermodynamic potential with appropriate 
constraints [6, 14]. The isotropic-nematic transition at $T^*=1.0$ with one order 
parameter is found to take place
at $\rho_{l}^{*}(=\rho_l\sigma_{0}^{3})=0.3325$ with change in density 
$\Delta \rho^* \left( \equiv(\rho_o-\rho_l)/\rho_l \right)=0.0086$ and 
order parameter $P_2=0.40$. If one uses the RY free energy functional then the
transition takes place at $\rho_{l}^{*}=0.3570$ with 
$\Delta \rho^*= 0.0055$ and $P_2=0.439$. The symmetry breaking part of PCFs
makes the isotropic phase unstable and induces the emergence of the ordered phase
at lower density. 
 
In Figs 1 and 2 we show some harmonic coefficients of DPCF in the director space for 
$T^*=1.0, \rho^*= 0.3361$ and $P_2=0.44$. While the harmonic coefficient $c_{2 2 0 0 0 0}(r^*)$
and $c_{2 2 0 1 -1 0}(r^*)$ shown in Fig 1 survive both in the isotropic ($P_2=0$)and 
the nematic phases($P_2\neq 0$),
the harmonic coefficients $c_{2 0 0 0 0 0}(r^*)$ and $c_{0 0 2 0 0 0}(r^*)$ survive only in
the nematic phase and vanish in the isotropic phase. The contribution arising due to symmetry 
breaking to the harmonic coefficients $c_{2 2 0 0 0 0}(r^*)$ and $c_{2 2 0 1 -1 0}(r^*)$
are shown by dot-dashed line in Fig 1 and are found to be small compared to the symmetry 
conserving part. Few selected harmonic coefficients of $h$ are shown in Figs 3 and 4
in director space.  In Fig 3 we
plot the harmonic coefficients $h_{2 0 0 0 0 0}(r^*)$ and $h_{0 0 2 0 0 0}(r^*)$ 
which survive only in the nematic phase and note the 
oscillatory behavior which continues to survive for large values
of the intermolecular separation $r^*(\equiv r/\sigma_0)$. In Fig 4 we plot
harmonic coefficient $h_{2 2 0 1 -1 0}(r^*)$ which is of fundamental importance
and defines nematic elastic constants[1]. As shown in the figure it decays as $1/r^*$ at
large distance. This long-range tail behaviour is attributed to the director transverse
fluctuations[4]. This behaviour has been shown analytically by Holovko and Sokolovska[4] 
and by computer simulation by Phoung and Schmid[2].  

The density functional approach
allows one to include more order parameters in the theory even though they are not included
in calculating the PCFs. This is done through the parametrization of $\rho({\bf x})$(see [6]). When we
take two order parameters $P_2$ and $P_4$ and use the free energy functional developed here
the transition is found to take place at $T^*=1$ with $\rho_{l}^{*}=0.317, \Delta\rho^*=0.026,
P_2=0.644$ and $P_4=0.332$. These values compare well with the computer simulation values,
$\rho_{l}^{*}=0.32, P_2=0.66$ and $P_4=0.29$[10]. This comparison suggests that the effect of order 
parameter $P_4$ on the PCFs is small. However, this does not mean that 
those harmonics that do not appear in the free energy functional are not sensitive
to the values of $P_4$. Inclusion of higher order parameters in calculation of PCFs 
is straightforward though computationally demanding.

In conclusion; we developed a method of solving the OZ equation with a closure relation
to get both the symmetry conserving and symmetry breaking parts of PCFs. Using these
correlation functions we constructed a free energy functional to study the freezing
transition and other properties of the ordered phase. Since the symmetry breaking 
parts of PCFs have features of the ordered phase including its geometrical 
packing, the free energy functional proposed here will allow us to study
various phenomena of the ordered phases including their relative stability[14]. 
Lastly we would like to emphasize that the theory developed here can easily be 
extended to study other ordered phases such as smectics and crystalline solids. 
  
We thank J. Ram for his help in computation. This work was supported by a research grant
from DST of Govt. of India, New Delhi.  

\begin{figure}[]
\includegraphics[height=6in,width=5in]{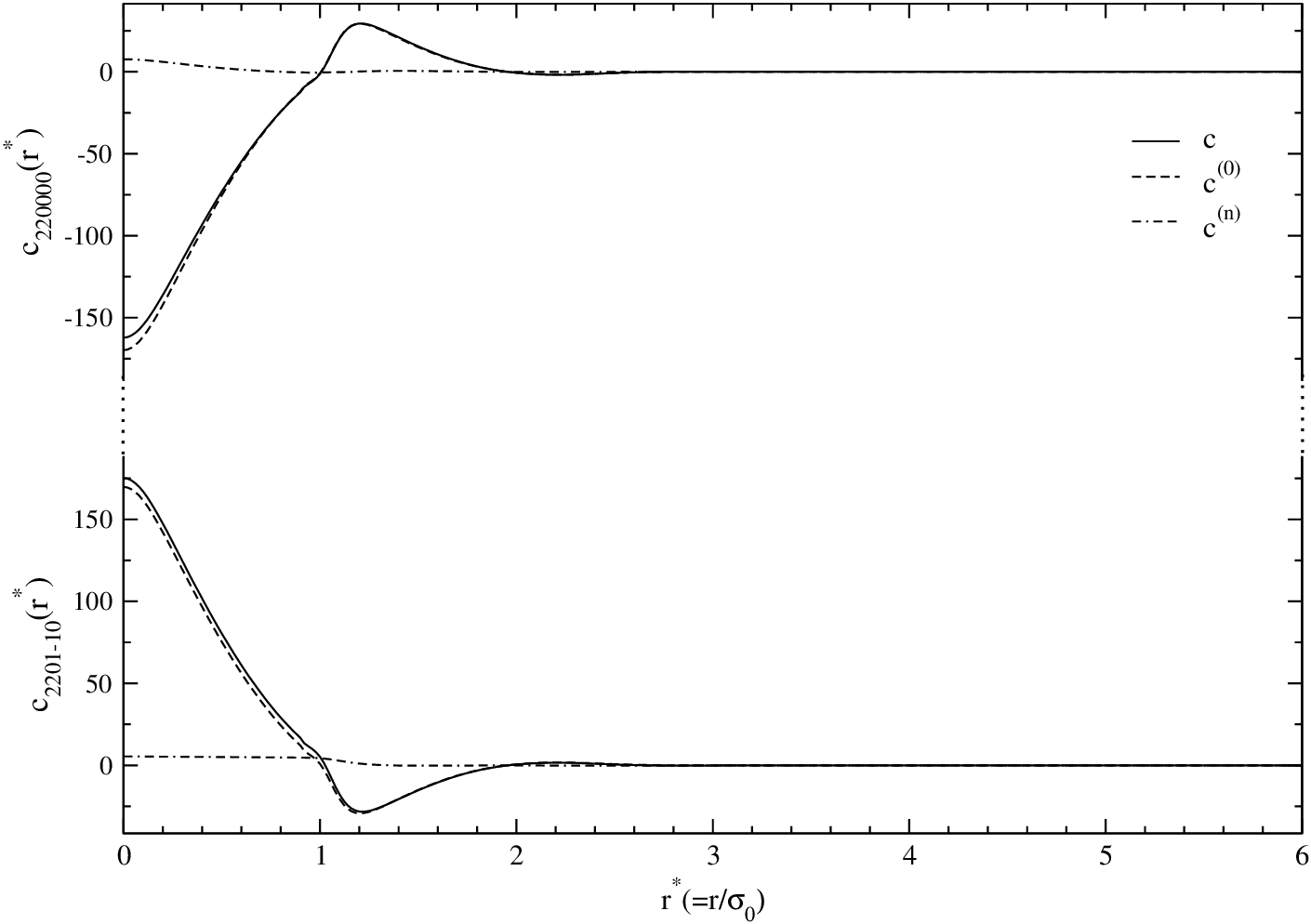}
\caption{ Harmonic coefficients of the DPCF in the director frame at 
$\rho^*=0.3361, T^*=1.0$ and $P_2=0.44$. The dashed line shows the contribution arising 
due to symmetry conserving part, dot-dashed line the symmetry broken part and the full line the
resultant.}
\end{figure}

\begin{figure}[]
\includegraphics[height=5in,width=5in]{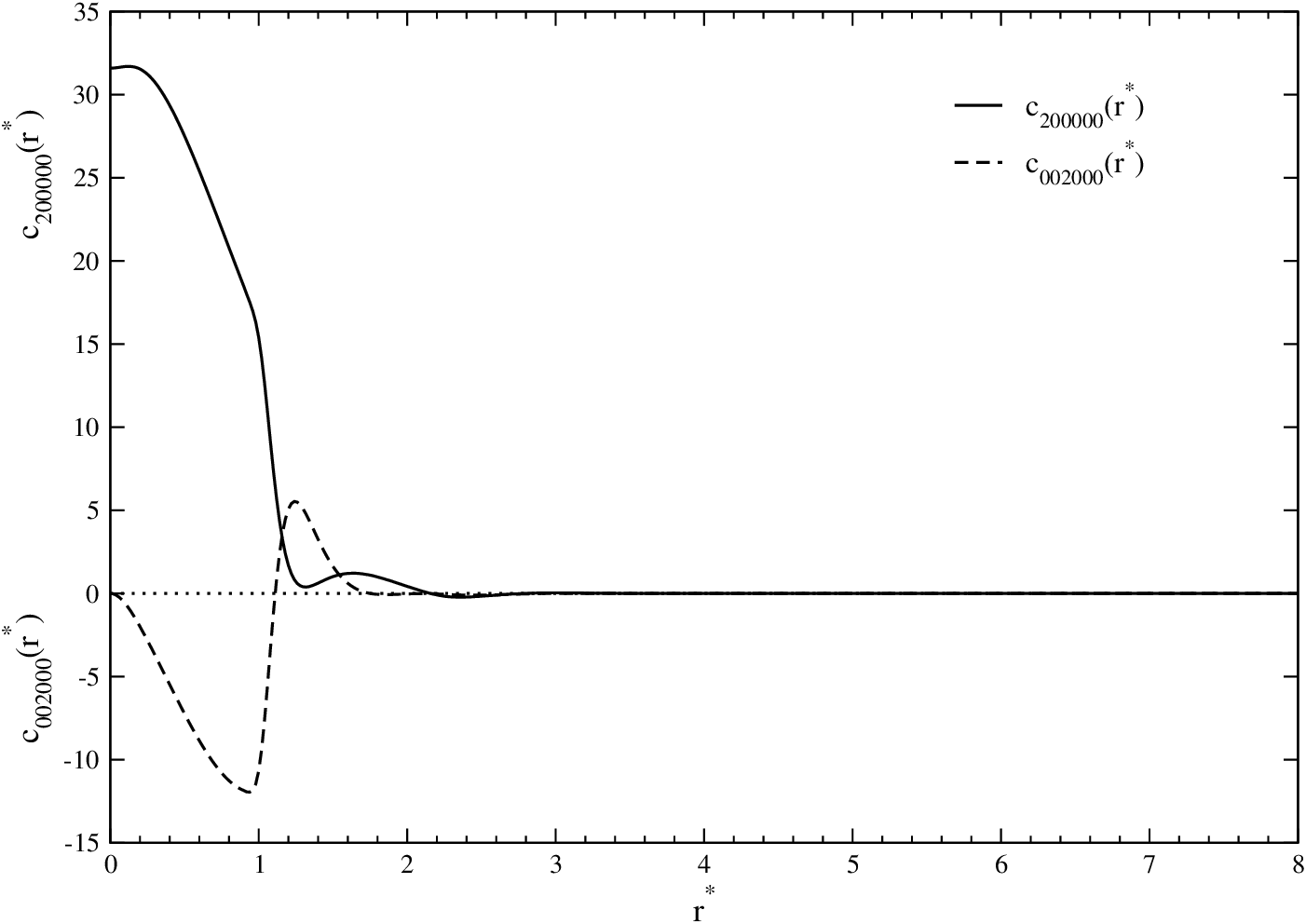}
\caption{Harmonic coefficients of the DPCF in the director frame at
$\rho^*=0.3361, T^*=1.0$ and $P_2=0.44$. The contributions to these coefficients arise due to the symmetry
breaking part only.}
\end{figure}

\begin{figure}[]
\includegraphics[height=5.0in, width=5.0in]{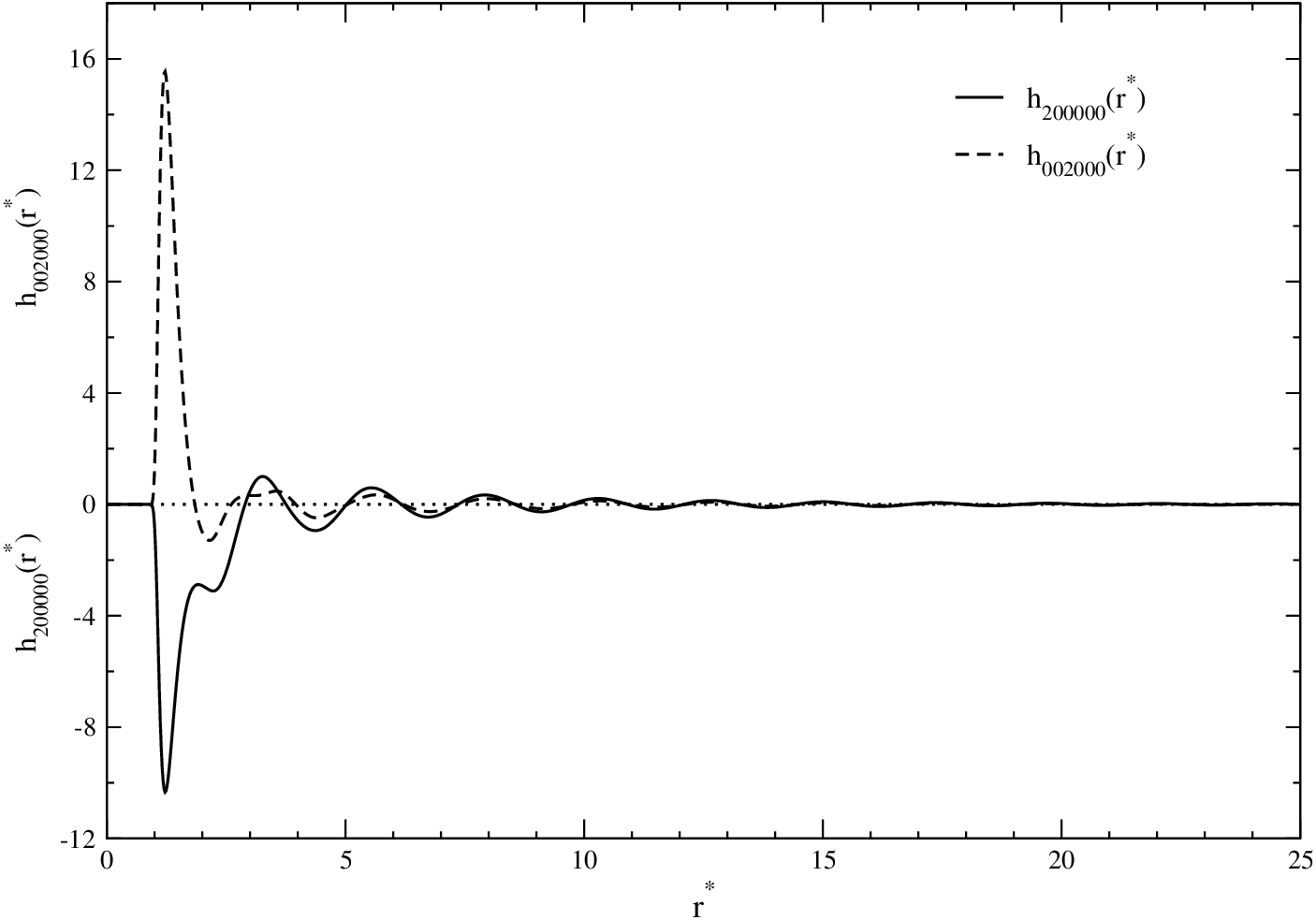}
\caption{Symmetry breaking h-harmonic coefficients in the director frame. 
 Details are same as in Fig 2.}
\end{figure}

\begin{figure}[]
\includegraphics[height=5in, width=5in]{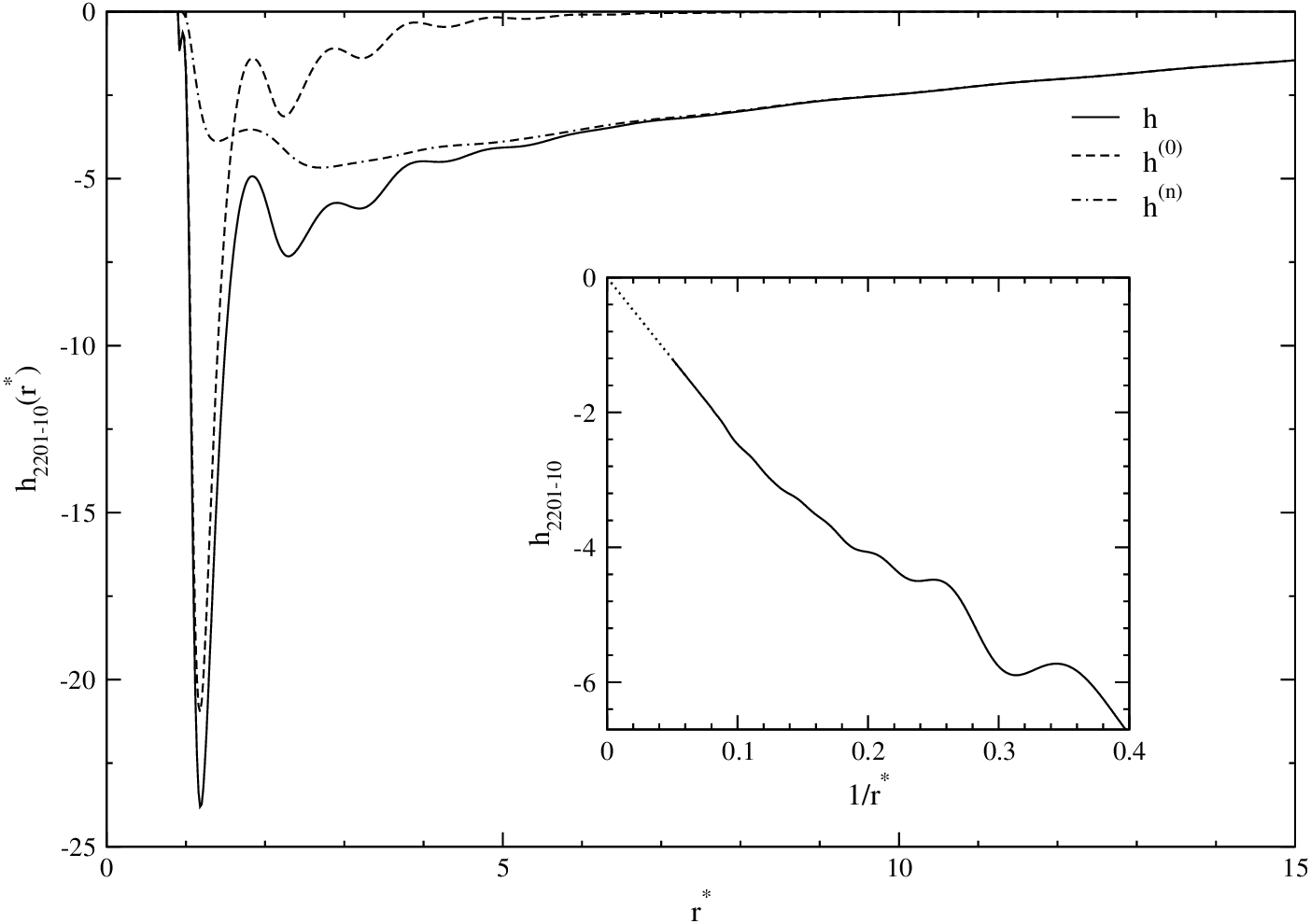}
\caption{ harmonic coefficient $h_{2201-10}$in the director frame at
$\rho^*=0.3361, T^*=1.0$ and $P_2=0.44$. Details are same as in Fig 1.
Inset shows the plot of $h_{2201-10}$ harmonics with respect to $1/r^*$, the
dotted line shows the extrapolated part.}
\end{figure}

\end{document}